\documentclass[preprint,12pt]{elsarticle}

\usepackage{amssymb}

\usepackage{lineno}

\biboptions{sort&compress}

\usepackage{xcolor}

\newcommand\beq{\begin{equation}}
\newcommand\beql[1]{\begin{equation} \label{#1}}
\newcommand\eeq{\end{equation}}
\newcommand\ben{\begin{eqnarray}}
\newcommand\een{\end{eqnarray}}
\newcommand\bea{\begin{array}}
\newcommand\eea{\end{array}}
\newcommand\bem{\begin{displaymath}}
\newcommand\eem{\end{displaymath}}

\newcommand\eqa[1]{Eq.(\ref{#1})}
\newcommand\eqb[1]{Eqs.(\ref{#1})}
\newcommand\eqc[1]{(\ref{#1})}

\newcommand\fig[1]{Fig.\ref{#1}}
\newcommand\figs[1]{Figs.\ref{#1}}
\newcommand\figg[1]{\ref{#1}}

\newcommand\sct[1]{Section~\ref{#1}}
\newcommand\scta[1]{Sections~\ref{#1}}
\newcommand\sctb[1]{~\ref{#1}}
\newcommand\app[1]{\ref{#1}}

\newcommand\qqa{\quad}

\newcommand\qqc{\qquad \qquad}

\newcommand\wse{\vspace{-3.5cm}}

\newcommand\Ssum[2]{\sum \limits_{#1}^{#2}}

\newcommand\dd[2]{\frac{{\rm d} #1}{{\rm d} #2}}

\newcommand\eV{ \, {\rm eV} }
\newcommand\MeV{ \, {\rm MeV} }

\newcommand\TeV{ \, {\rm TeV} }

\newcommand\EeV{ \, {\rm EeV} }

\newcommand\Xmax{X_{\rm max}}
\newcommand\AXmax{\langle X_{\rm max} \rangle}
\newcommand\SXmax{\sigma^{2}_{\rm max}}
\newcommand\sxmax{\sigma_{\rm max}}
\newcommand\ssxmax{\sigma (X_{\rm max})}

\newcommand\AXmaxA{\langle X_{\rm max} \mid A \rangle}
\newcommand\AAXmaxA{\langle \langle X_{\rm max} \mid A \rangle \rangle}
\newcommand\AXmaxAp{\langle X_{\rm max} \mid A=1 \rangle}
\newcommand\AXmaxAFe{\langle X_{\rm max} \mid A=56 \rangle}

\newcommand\Axmax{\langle x_{\rm max} \rangle}

\newcommand\AXmaxp{\langle \Xmax^{\rm p} \rangle} 
\newcommand\AXmaxFe{\langle \Xmax^{\rm Fe} \rangle}

\newcommand\SXmaxA{\sigma^{2} (X_{\rm max} \mid A)}
\newcommand\SSXmax{\sigma^{2} (X_{\rm max})}
\newcommand\SXmaxAA{\sigma^{2} (\langle X_{\rm max} \mid A \rangle)}
\newcommand\ASXmaxA{\langle \sigma^{2} (X_{\rm max} \mid A) \rangle}

\newcommand\AXmaxex{\langle X_{\rm max}^{\rm exp} \rangle} 
\newcommand\Dex{D^{\rm exp}}

\newcommand\sfr{\sigma_{\rm fr}^{2}}
\newcommand\ssh{\sigma_{\rm sh}^{2}}
\newcommand\saa{\sigma_{\ln {\rm A}}^{2}}
\newcommand\sfrA{\langle \sigma_{\rm fr}^{2} \rangle}
\newcommand\sshA{\langle \sigma_{\rm sh}^{2} \rangle}
\newcommand\sfro{\sigma_{\rm fr,0}^{2}}
\newcommand\ssho{\sigma_{\rm sh,0}^{2}}
\newcommand\sfroo{\sigma_{\rm fr,0}}
\newcommand\sshoo{\sigma_{\rm sh,0}}

\newcommand\An{A_{\rm n}}
\newcommand\Ax{A_{\rm x}}

\newcommand\pa{p_{\rm A}}
\newcommand\lnAA{\langle \ln A \rangle}
\newcommand\lnAtA{\langle \ln^{2} A \rangle}
\newcommand\lnAs{\sigma_{\rm \ln A}^{2}}

\newcommand\gIcmS{ \, {\rm gcm^{-2}} }
\newcommand\mbb{ \, {\rm mb} }

\newcommand\Log{{\rm Log}}
\newcommand\ovl[1]{\overline{#1}}

\newcommand\smu{\sigma_{\rm M}^{2}}
\newcommand\ska{\sigma_{\rm \kappa}^{2}}
\newcommand\smuo{\sigma_{\rm M,0}^{2}}
\newcommand\skao{\sigma_{\rm \kappa,0}^{2}}

\journal{Astroparticle Physics}

\begin{document}

\begin{frontmatter}

\title{Maximum entropy analysis of cosmic ray composition}


\author[nos01]{Dalibor Nosek}
\ead{nosek@ipnp.troja.mff.cuni.cz}
\author[ebr01]{Jan Ebr}
\author[ebr01]{Jakub V\'{i}cha}
\author[ebr01]{Petr Tr\'avn\'{i}\v{c}ek}
\author[nos02]{Jana Noskov\'a}
\address[nos01]{Charles University, Faculty of Mathematics and Physics,
Prague, Czech Republic}
\address[ebr01]{Institute of Physics, 
Academy of Sciences of the Czech Republic, 
Prague, Czech Republic}
\address[nos02]{Czech Technical University, Faculty of Civil Engineering,
Prague, Czech Republic}

\begin{abstract}
We focus on the primary composition of cosmic rays with the highest 
energies that cause extensive air showers in the Earth's atmosphere. 
A way of examining the two lowest order moments of the sample 
distribution of the depth of shower maximum is presented. 
The aim is to show that useful information about
the composition of the primary beam can be inferred with limited 
knowledge we have about processes underlying these observations.
In order to describe how the moments of the depth of shower maximum 
depend on the type of primary particles and their energies, 
we utilize a superposition model.
Using the principle of maximum entropy, we are able to determine what 
trends in the primary composition are consistent with the input data, 
while relying on a limited amount of information from shower
physics.
Some capabilities and limitations of the proposed method are 
discussed. 
In order to achieve a realistic description of the primary mass 
composition, we pay special attention to the choice of the parameters 
of the superposition model.
We present two examples that demonstrate what consequences can be
drawn for energy dependent changes in the primary composition.
\end{abstract}

\begin{keyword}
Ultra--high energy cosmic rays \sep
Extensive air showers \sep
Cosmic ray composition
\end{keyword}

\end{frontmatter}


\section{Introduction}
\label{Sec01}

The mass composition of cosmic rays (CR) is an important issue 
in astroparticle physics research. 
The energy dependence of the primary mass distribution can provide 
useful information about ultra--high energy cosmic rays (UHECR) origin, 
their acceleration mechanisms and propagation through the galactic 
and extragalactic space. 
The mass observables can help to understand typical spectral features 
of UHECRs, the ankle observed at about $4~\EeV$ and the steep flux 
suppression at energies above $30~\EeV$.
In addition, the knowledge of the mass composition of UHECRs allows 
for an easier search for their sources or even investigation 
of basic characteristics of these sources.

In seeking for the masses of primary UHECR particles, 
the longitudinal development of extensive air showers (EAS) 
of secondary particles created in the Earth's atmosphere is 
usually examined. 
The penetration depth at which the CR shower reaches the maximum 
number of particles, $\Xmax$, reflects the type of the primary 
particle causing this shower.
The average depth of shower maximum for a set of CR showers 
detected at a given energy range, $\AXmax$, and its standard 
deviation, $\sxmax = \ssxmax$, are then used to describe 
the main features of the primary mass composition.
The quantitative interpretation of these data in terms of 
primary mass demands an accurate model of hadronic 
interactions.
Usually, particles having a mass ranging from protons to 
iron nuclei are considered as responsible for the observed 
shower profiles of CR events.
However, there is little information from the theory of 
what UHECR species are registered in current large CR detectors.
Since the relevant phase space regions have not 
been explored in laboratory experiments, required interaction 
parameters are extrapolated from lower energy experiments,
making the composition analysis uncertain.

Recent results from the Pierre Auger Observatory indicate 
a mixed CR composition with a transition from light to heavier 
primaries at the ankle region~\cite{Aug01,Aug02,Aug03,Aug04}.
Measurements of $\AXmax$ show a flattening of the elongation rate 
near above $2 \EeV$. 
In addition, fluctuations of $\Xmax$ expressed by the standard 
deviation $\sxmax$ were found to decrease from 
approximately $60 \gIcmS$ at $2 \EeV$ to about $30 \gIcmS$ 
at $40 \EeV$. 
However, no such trends were observed by the HiRes and Telescope Array 
experiments.
Their analyses prefer light primaries at the highest 
energies~\cite{Hir01,Tar01}.
But it is not excluded that the observed inconsistencies may be due to
the fact that detector effects were not fully eliminated.
All the current experiments agree in suggesting a light CR composition 
below $2 \EeV$ at the level of their systematic 
uncertainties~\cite{Aug01,Aug02,Aug03,Aug04,Hir01,Tar01,Yak01}.

Problems related to the mass composition of UHECRs have been 
widely communicated in the literature, see e.g. Refs.~\cite{Ung01,Bar01} 
for recent reviews.
The energy dependencies of the average logarithmic mass and of 
its standard deviation, as measured by the Pierre Auger Observatory, 
have been recently examined under the assumption of selected 
models of hadronic interactions~\cite{Aug04,Aug05,Aug06,Aug07}.
Other methods based on a given parametrization of the distribution 
of the depth of shower maximum for the study of the mass composition 
of UHECRs have been introduced, among others see e.g. 
Refs.~\cite{Dom01,Rig01}.
The need of the muon number measurement is often emphasized 
for estimating the spread of masses in the beam of primary UHECR 
species~\cite{Aug08,Aug09,Aug10,Aug11,You01}.
However, recently analyzed data from the Pierre Auger Observatory 
indicates that the muonic component of air showers is not well 
described by the current models of hadronic interactions 
used for EAS simulations~\cite{Aug12,Aug13}.
Also, different statistical tools have been used to obtain 
information about the primary mass composition on 
an event--by--event basis, see e.g. Refs.~\cite{Amb01,Rig02,Aug14}.
Finally, it is worth noting that the knowledge of the chemical 
composition of UHECRs was shown to play a crucial role in studies 
trying to describe the anisotropy signal and, consequently, to estimate 
properties of CR sources~\cite{Lem01,Aug15,Tay01,Tay02,Tay03}.

In case of experiments that measure the depths of shower 
maximum, the distribution of the mass of primary particles can be
inferred only with the help of sets of simulated reference 
showers. 
However, since the shower properties are not yet fully understood, 
currently available models of hadronic interactions provide  
different solutions to the composition problem, for the recent
analysis of the Auger data see Ref.~\cite{Aug07}.
Dealing with the mean and variance of the depth of shower maximum, 
mass observables of primary particles are usually estimated and, 
eventually, their relationship is examined using hadronic interaction 
models~\cite{Ung01,Bar01,Aug05}.
The power of this combined analysis has been repeatedly 
emphasized.
Nevertheless, the predictions of existing models are different and 
in some cases even indicate possible inconsistencies in the modeling 
of hadronic interactions~\cite{Aug04,Aug05}.

Inspired by these findings, we examine what can actually be 
obtained using just a set of the two lowest order $\Xmax$ moments.
We address the issue of how to assess what trends in the primary
composition are most strongly supported by this data if only 
a limited knowledge of EAS properties is available. 
The proposed method and the interpretation of its results is quite 
distinct from and independent of other more conventional procedures 
used in composition studies.
In particular, our inference procedure is designed to exploit 
incomplete information about investigated phenomena and provides 
their probabilistic interpretation.

With the aim to deduce relative occupancy of primary particles, 
we relate shower observables to average masses of incident primaries 
using an air shower model.
We intentionally made an attempt to account for the basic properties 
of the longitudinal EAS development independently of the assumptions 
about detailed features of hadronic interactions.
Instead, we used the fact that the current data and its subsequent 
analysis, when faced with air shower simulations, are not able 
to undermine the validity of a simple superposition ansatz~\cite{Ung01}.
This choice allows us to classify obtained solutions in the space 
of physically reasonable parameters.
Moreover, it enables us to assess the properties of different models 
of hadronic interactions. 
Finally, when we present the resultant primary composition, 
existing knowledge about EAS physics is considered.
In our treatment, the superposition model was supplemented by simple 
considerations providing us other mass dependent terms relevant for 
EAS physics~\cite{Ung01}.
As a result, the two lowest order $\Xmax$ moments were parametrized 
in a similar manner as originally suggested 
in Refs.~\cite{Lin01,Lin02}.

We focused on how to gain credible information on the primary mass 
composition that takes account of our incomplete knowledge 
of the underlying processes leading to the observation
of the two lowest order $\Xmax$ moments.
For this purpose, we adopted the principle of maximum 
entropy~\cite{Jay01,Jay02,Rao01,Pre01}. 
This criterion, without any other assumptions about $\Xmax$
data, allows us to choose a unique well--behaved solution 
among various options how to combine primary components so as 
to obtain the two lowest order sample moments.

The method of maximum entropy relies on the properties of 
entropy as a measure of uncertainty.
It sets the task to return a maximally noncommittal distribution 
of a quantity that is constrained by information obtained in experiment.
It is worth stressing that such a solution does not necessarily 
provide an unambiguous description of the process that generates 
the observed data.
Instead, this method provides us with the probability distribution 
of the underlying quantity which is most strongly supported by the facts
while using as little additional information as possible in order to
avoid unintentionally assuming more than is really known.
This scheme is not only backed by a compelling statistical motivation, 
but also fairly simple to implement, yet sufficiently general. 
It is widely used in many branches of science, for recent review 
of its basic ingredients, aspects and applications see e.g. 
Ref.~\cite{Pre01}.

In the context of composition studies, the proposed method 
treats the two lowest order $\Xmax$ moments, and possibly other 
average shower observables, on an equal footing.
Having these moments, the probabilities are uniquely assigned 
to selected primary particles that are assumed to cause observed 
air showers. 
For this, we need a specific shower model that converts shower 
observables into the mass number space.
The resultant distribution of the incident particles is then 
obtained from the available data without any further assumptions 
about the properties of this data. 
Such a solution enables us to draw minimally biased conclusions 
about the composition of the beam of primary particles within 
the framework of a chosen shower model.
More importantly, we can find a set of acceptable solutions 
with maximum entropy in the parameter space of the shower model 
and check whether the available models of hadronic interactions 
can provide such solutions. 
The analogous interpretation of mass composition measurements 
does not seem to have been previously documented.

The structure of the paper is as follows. 
In~\sct{Sec02}, the air shower model is introduced 
supplemented by~\app{App01}. 
Particular attention is paid to the choice of model parameters.
The original contribution of our study, the inference procedure 
for the composition determination, is described in~\sct{Sec03}.
In this section, we present a way to use the partition method and 
point out the probabilistic interpretation of its output.
The essential features of the underlying principle of maximum entropy 
are summarized in~\app{App02}.
Examples are presented and discussed in~\sct{Sec04}.
The paper is concluded in~\sct{Sec05}.

\section{Air shower model}
\label{Sec02}

Let us assume that a depth of shower maximum $\Xmax$ is observed
when a UHECR particle with mass $A$ and energy $E$ hits the upper 
part of the Earth's atmosphere. 
We treat the former two quantities as dependent random variables, 
$\Xmax = \Xmax(A)$. 
The primary energy is considered to be a known parameter.
For the longitudinal shower development we utilize the superposition 
model in which a primary nucleus is regarded as a superposition 
of $A$ nucleons of energy $E/A$. 
We assume that the mean depth of shower maximum of a set of showers 
caused by the same primaries is a linear function of the decimal
logarithm of their energies per nucleon~\cite{Ung01}
\beql{E01}
\AXmaxA = C + D~\Log \left( \frac{E}{E_{0} A} \right).
\eeq
Here, $C = \Axmax(E_{0})$ denotes the mean depth of shower maximum 
for protons with a reference energy of $E_{0}$, 
$D = \dd{\Axmax}{\Log E}$ is the proton elongation rate and 
the proton mean depth of shower maximum is denoted by
$\Axmax = \Axmax(E) = \AXmaxAp$.
The model parameters $C$ and $D$ depend on the properties of hadronic 
interactions.\footnote{
Instead of the parameters $C$ and $D$, usually the mean values of 
the depth of shower maximum for primary protons and iron nuclei 
at a chosen energy are used, i.e. $\AXmaxp$ and $\AXmaxFe$.
In our notation scheme, 
$\AXmaxp = \Axmax(E_{0}) = C$ for protons and, in a similar manner, 
$\AXmaxFe = \AXmaxAFe(E_{0}) = C - D~\Log(56)$
for iron nuclei.
}
Although weak dependence of the parameter $D$ on the primary proton 
energy is expected~\cite{Ung01}, we consider it to be constant.

In a similar way, the conditional variance of the depth 
of shower maximum of a set of showers initiated by the same 
primaries with equal energies consists of two terms~\cite{Ung01}, 
namely,
\beql{E02}
\SXmaxA = \sfr + \ssh.
\eeq
Here, $\sfr = \sfr(A,E)$ is the variance of the depth where 
the first interactions of the primary particles take place.
The variance of the depth of shower maximum associated 
with the subsequent shower development is denoted by 
$\ssh = \ssh(A,E)$.

Hence, for a mixed beam of primaries with a given energy, the total mean 
and total variance of the depth of shower maximum that are to be confronted 
with measurements are, respectively,
\beql{E03}
\AXmax = \AAXmaxA = \Axmax - d \lnAA,
\eeq
and
\beql{E04}
\SXmax = \SSXmax = \sfrA + \sshA + d^{2} \saa,
\eeq
where $D = d \ln 10$ was inserted and the law of total variance 
was used, i.e. 
$\SXmax = \ASXmaxA + \SXmaxAA$ (see e.g. Ref.~\cite{Rao01}).
Except for $\Axmax$, the other mean values on the right 
hand sides in~\eqb{E03} and~\eqc{E04}, and the variance 
on the right hand side in~\eqa{E04} as well, are calculated 
over the mass numbers of primary particles.
The mean and variance of their logarithmic mass are denoted  
by $\lnAA$ and $\saa$, respectively.
More detailed information about these simple dependencies 
and their confrontation with experimental data can be found, 
for example, in Ref.~\cite{Ung01}.

The mean and variance of the depth of shower maximum are directly 
connected with the distribution of the logarithmic masses of primary 
particles causing studied showers. 
For a given primary energy, the sample mean $\AXmax$ informs us about 
the average value of the mass distribution of primaries hitting the upper 
edge of the Earth's atmosphere.
Apart from mass dependent fluctuations in shower development, 
the sample variance $\SXmax$ carries information about the spread 
of the mass distribution of primaries as they were created 
in CR sources and eventually modified during their propagation 
through the space. 

Let us remind here that in case of experiments that measure 
$\AXmax$ the average logarithmic mass of primary CR particles, 
$\lnAA$, is usually derived from~\eqa{E03} 
with the parameters inferred from predictions based on 
a model of hadronic interactions, see e.g.~Ref.~\cite{Ung01}.
Moreover, if experimental values of $\SXmax$ 
are available and if shower fluctuations are determined 
based on current predictions, the sample variance of 
the logarithmic mass of primary particles, $\saa$, can 
be estimated from~\eqa{E04}. 
The importance of such a kind of a combined analysis to identify 
the primary mass composition using mass dependent shower observables 
has been repeatedly emphasized~\cite{Lin01,Lin02,Ung01,Aug05}.
The compatibility of both measurements with shower simulations 
can be judged within the $\lnAs$--$\lnAA$ plane~\cite{Aug05} or 
in the $\sxmax$--$\AXmax$ plane~\cite{Ung01}.
In particular, notwithstanding the limitation imposed by the experimental 
resolution, the Auger data combined with the current model 
predictions suggests a change from light to medium light 
composition with a minimum in the average logarithmic mass 
$\lnAA$ below $3 \EeV$~\cite{Aug04,Aug05}. 
More importantly, the logarithmic mass spread $\lnAs < 2$
was found for most models in the whole energy range 
indicating that mostly neighboring nuclei are mixed~\cite{Aug04,Aug05}.

Here, using the same information, $\AXmax$ and $\sxmax$, 
we address a different problem. 
We focus on the determination of the probability distribution 
for primary species.  
In order to obtain detailed information about this distribution, 
we assess the usability of the simple model for the EAS development 
given in~\eqb{E03} and~\eqc{E04}.
In evaluating the partition probabilities of primary particles, 
we encountered the problem of how to limit the parameter space 
of the shower model. 

In our approach, the mean depth of shower maximum for primary 
protons, $\Axmax$, is determined by the two energy independent 
parameters, by the mean depth of shower maximum at a reference 
energy $E_{0}$, $C = \Axmax(E_{0})$, and by the elongation rate 
$D = d \ln 10$, see~\eqa{E01}. 
They depend on the properties of hadronic interactions
and can be determined within a chosen model.
Here, we proceed differently.
We examine these parameters, while keeping 
the fluctuations parameters in acceptable ranges.
We search for solutions across the $C$--$D$ plane to document 
the conditions under which models of hadronic interactions can provide 
a maximally unbiased description of experimental information.

Therefore, in the first step of the analysis, we leave both 
parameters $C$ and $D$ free having values in a reasonable 
domain subject to limiting conditions that are dictated by 
the solvability of the decomposition problem.
Assuming that primary particles with the mass numbers 
$A \in \langle \An, \Ax \rangle$ are responsible for 
the two lowest order $\Xmax$ moments, then, among other 
constraints, for the parameters $C$ and $D$ one gets 
from~\eqa{E03}
\beql{E05}
d \ln \An \le \Axmax - \AXmax \le d \ln \Ax.
\eeq
These inequalities are assumed to be valid for all $\AXmax$ values 
throughout the whole energy range considered.

In the second step, when a particular solution with maximum 
entropy is described, we emphasize the need for further information 
about what kind of air showers are caused by various primaries. 
For this purpose, we supplement our analysis with 
the current knowledge about EAS physics.
Our treatment requires only mean properties of the depth of shower 
maximum for protons at a reference energy and their evolution with 
mass and energy, see~\eqb{E03} and~\eqc{E04}, and~\app{App01}.
Specifically, we use the superposition model with parameters, 
the ranges of which allow us to describe the results of 
measurements~\cite{Aug01,Aug02,Aug03,Aug04,Hir01,Tar01,Yak01}, 
conditioned by outputs of the existing models of hadronic 
interactions~\cite{Wer01,Wer02,Ost01,Ahn01}.

First, we note that the average value of the depth of shower 
maximum $\AXmaxex \approx 770 \gIcmS$ and the elongation rate 
of $\Dex \approx 30 \gIcmS$ were deduced from experiments at energy 
of $10 \EeV$ where probably a mixed primary composition is 
observed~\cite{Aug04,Ung01,Bar01,Aug07}.
Experimentally, the value of the elongation rate seems to increase 
with the decreasing energy,
giving about $C \approx 730 \gIcmS$ and $D \approx 50 \gIcmS$, 
or even larger values, at a reference energy of $E_{0} = 1~\EeV$
if the primary mass composition dominated by protons is assumed.
These findings indicate that acceptable values of the
relevant parameters, respectively, are to be chosen in the vicinity 
or above the indicated values, whatever the composition of the primary 
beam at the reference energy.

In practical applications, we prefer to use such a set of 
parameters that is not in conflict with the current model predictions 
(see e.g. Ref.~\cite{Ung01} and references therein).
For this, we fitted the parameters of the superposition 
model for different models of hadronic interactions, 
for H, He, N and Fe primary nuclei. 
For each model, about $10^{4}$ CONEX showers~\cite{Ber01} with 
primary energies between $0.1 \EeV$ and $10 \EeV$ were generated. 
We obtained 
$(C,D) \approx (752, 56), (739, 53)$ and $(738, 54) \gIcmS$ with 
uncertainties below $3 \gIcmS$ 
for the EPOS-LHC~\cite{Wer01,Wer02}, QGSJet~II-04~\cite{Ost01} 
and Sibyll~2.1~\cite{Ahn01} models, respectively. 
The superposition predictions for $\AXmax = \AXmax(E)$ derived from 
these considerations are shown in~\fig{F01}.

Based on all these arguments, in the two examples 
presented in~\sct{Sec04} we describe in detail only those solutions 
that meet the conditions 
$C \in \langle 720, 750 \rangle \gIcmS$ and 
$D \in \langle 50, 65 \rangle \gIcmS$.
In this sense, resultant probability distributions for incident 
particles, as presented in the examples, depend on the predictions 
of models of hadronic interactions.

For a given type of primary particles with a given energy, 
fluctuations in the depth of shower maximum are related 
to the dispersion of the depth of the first interaction, 
$\sfr = \sfr(A,E)$, and to the stochastic nature of secondary 
interactions occurring along the shower development, 
$\ssh = \ssh(A,E)$, see also~\eqa{E02}.
In order to estimate these variations, we used a simple 
phenomenological approach described in~\app{App01}.
The variance induced in the first or main interaction, 
$\sfr = \sfr(A,E)$, was deduced from the measured 
$p$--air cross section~\cite{Aug16} and its extrapolated energy 
dependence accompanied by simple geometrical and 
statistical considerations.
Except for systematic effects, this variance depends also 
on the assumptions about hadronic interactions~\cite{Aug16}.
The properties of the variance associated with the shower development, 
$\ssh = \ssh(A,E)$, were inferred using a simple concept of 
multiplicity and elasticity of hadronic interactions~\cite{Ung01}.

The fluctuation parameters for showers induced by protons can 
be constraint using the experimental data~\cite{Aug01,Aug02,Aug03,Aug04}, 
see~\app{App01}.
For our purpose, we also estimated them with the help of the current models 
of hadronic interactions.
We fitted relevant parameters using about $10^{4}$ CONEX 
showers~\cite{Ber01} that were generated with energies around 
$1 \EeV$ for H, He, N and Fe primary nuclei.
For proton showers at $1 \EeV$ we obtained $\sxmax \approx 62, 65$ 
and $57 \gIcmS$ with uncertainties below $3 \gIcmS$ 
using the EPOS-LHC~\cite{Wer01,Wer02}, QGSJet~II-04~\cite{Ost01} 
and Sibyll~2.1~\cite{Ahn01} models, respectively. 
Following the approach described in~\app{App01}, the shower fluctuations 
of the first interaction and of the subsequent shower development 
were separated in experimentally supportable proportion 
$\frac{\sfro}{\ssho} \approx \left( \frac{46}{38} \right)^{2}$.
With such parameters we calculated the energy and mass dependent 
estimates for the shower variances which apply under the method 
described in~\app{App01}.
These estimates of $\sxmax = \sxmax(E)$ are shown in~\fig{F02}.

Finally, it is worth noting that characteristics for the shower development 
required in the analysis
($\AXmax = \AXmax(E,A)$ and $\sxmax = \sxmax(E,A)$)
can be directly
calculated in cascade simulations utilizing various models of hadronic 
interactions.
Consequently, some of the assumptions for these interactions 
can be checked for inconsistencies if sufficiently accurate data 
is available.
However, such an analysis based on Monte Carlo estimates of shower 
characteristics carried out with a set of currently known data 
is beyond the scope of this study.

\section{Partition problem}
\label{Sec03}

First suppose that all the parameters of the shower model 
are fixed.
Given the two lowest order $\Xmax$ moments, \eqb{E03} 
and~\eqc{E04} represent the system of two linear equations 
for a set of unknown fractions of primary particles 
at a given energy. 
In the case of a two--component partition, this system is 
overdetermined and its solution is not guaranteed even if 
both equations are dependent since the sum of the fractions 
is normalized to one.
In a similar manner, a three--component solution
satisfying~\eqb{E03} and~\eqc{E04} at a given energy may or 
may not exist depending on the model parameters.
If we reverse this problem, leaving the values of some model 
parameter free, then the deterministic solubility of 
the system of two equations with given $\Xmax$ moments can 
serve as a filter for a reasonable set of the model parameters.

When the number of unknown fractions is larger, for four or 
more components, the system of~\eqb{E03} and~\eqc{E04} 
is underdetermined and its unambiguous solution, if it exists, 
requires additional conditions.
For this purpose, we adopted the maximum entropy 
principle~\cite{Jay01,Jay02,Rao01,Pre01}. 
In this scheme, the probability distribution of primary 
particles is deduced from given data simply by maximizing 
missing information, for more details see~\app{App02}.

We treat each individual shower as a random phenomenon  
in the sense that its properties are given by choosing randomly 
the mass of a primary particle initiating its development. 
We further assume that the depth at which the subsequent cascade 
of particles reaches its maximum is inferred from the longitudinal 
profile of the shower.
Then, the distribution of the depth of shower maximum is constructed 
for events in a selected energy range and its lowest order moments, 
$\AXmax$ and $\SXmax$, are determined.

In order to obtain reliable information about the distribution 
of primary particles from these sample moments, we choose 
two independent constraints, respectively,
\beql{E06} 
F_{1}(A) = d \ln A,
\eeq 
and 
\beql{E07}
F_{2}(A) = d^{2} \ln^{2} A + \sfr + \ssh.
\eeq
The average values of both these constrains taken over the mass 
numbers of primary particles are related to the two lowest order moments 
of the logarithmic mass distribution, $\lnAA$ and $\lnAs$.
Using \eqb{E03} and \eqc{E04}, we have 
\beql{E08}
\langle F_{1} \rangle = 
d \lnAA = 
\Axmax - \AXmax,
\eeq
and
\beql{E09}
\langle F_{2} \rangle =
d^{2} \lnAtA + \sfrA + \sshA = 
\SXmax  + \left( \Axmax - \AXmax \right)^{2}.
\eeq
Except for the model value of the mean depth of shower maximum for
protons, $\Axmax$, 
the average values of the two constraints are given by the available 
information contained in the $\Xmax$ measurements.

In our analysis, the probability distribution of primary particles 
at a given energy is dictated by the maximum entropy principle as 
described in~\app{App02}. 
Knowing the total sample mean and variance, $\AXmax$ and $\SXmax$, 
and adopting the model value of $\Axmax = \Axmax(C,D)$,
the shape of this distribution is determined by~\eqa{B03}.
The corresponding two Lagrange multipliers are derived numerically 
in such a way that the average values of the two constraints written 
in~\eqb{E08} and~\eqc{E09} are satisfied within the framework of 
the shower model described in~\sct{Sec02} and~\app{App01}.

It is worth first pointing out the probabilistic nature 
of the maximum entropy analysis. 
Solving the partition problem we do not claim that the composition 
of primary particles initiating studied EAS is unequivocally given 
by the deduced probability distribution.
Instead, we state that given the experimental data, we gain a consistent 
description of the underlying processes within the model of the EAS 
development which deliberately avoids assuming any other facts. 

In order to prevent possible misinterpretation, we stress that 
the maximum entropy method has nothing to do with likelihood 
or Bayesian estimates. 
Whereas these two methods deduce fractions of primaries and 
their uncertainties from distributions of data, we do the reverse, 
deducing the probability distribution for primaries from 
a limited set of observable characteristics.
In the search for the most accurate solutions, additional 
assumptions about the data are necessary in the former cases 
(e.g. likelihood function).
In the latter case, one admits the most ignorance beyond 
the prior data in order to obtain the least distorted information.

The proposed analysis scheme provides distributions of primary species 
that are consistent with the sample mean values and variances of 
the depth of shower maximum. 
When the underlying EAS physics is known, this method can exploit
any other set of input information about shower observables 
(for example, muon numbers, their production depths or signal rise 
times etc.), and constrain thus the space of physically admissible 
solutions. 

\section{Two examples}
\label{Sec04}

Hereafter we demonstrate that the partition method supplemented by 
the air shower model provides a reliable tool for estimating 
the spread of masses of primary particles causing cascades 
of secondaries in the Earth's atmosphere.
The two lowest order moments of the distribution of the depth 
of shower maximum were used as the input. 
We examined their evolution with primary energy.
The mass composition of primary particles was derived using 
the partition method described in~\sct{Sec03}.
Results were interpreted within the shower model presented 
in~\sct{Sec02}. 
\begin{figure}[ht!]
\wse
\includegraphics*[width=0.99\linewidth]{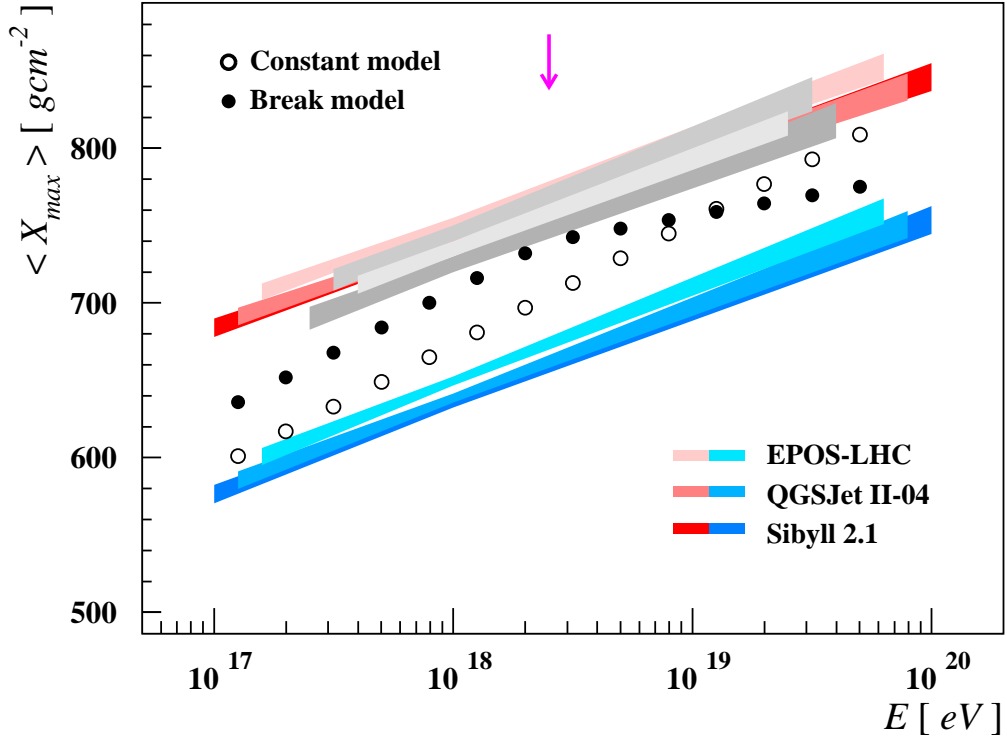}
\caption{\small
Average values of the depth of shower maximum are shown 
as functions of primary energy.
Circles are for the constant elongation rate (constant model).
Black full points are for the break model with two different
elongation rates as indicated by a magenta arrow.  
Bands with red (blue) shade show $\AXmax$ predictions 
of the superposition model for primary protons (iron nuclei) 
that we received using CONEX showers generated with primary 
energies around $1 \EeV$ within the indicated models of hadronic 
interactions.
Gray bands show $\Axmax(E)$ for protons that we chose  
in seeking for four--component solutions of the constant model 
(light middle band) and break model (two darker gray bands), 
see~\scta{Sec04a} and~\sctb{Sec04b}.
}
\label{F01}
\end{figure}
\begin{figure}[ht!]
\wse
\includegraphics[width=0.99\linewidth]{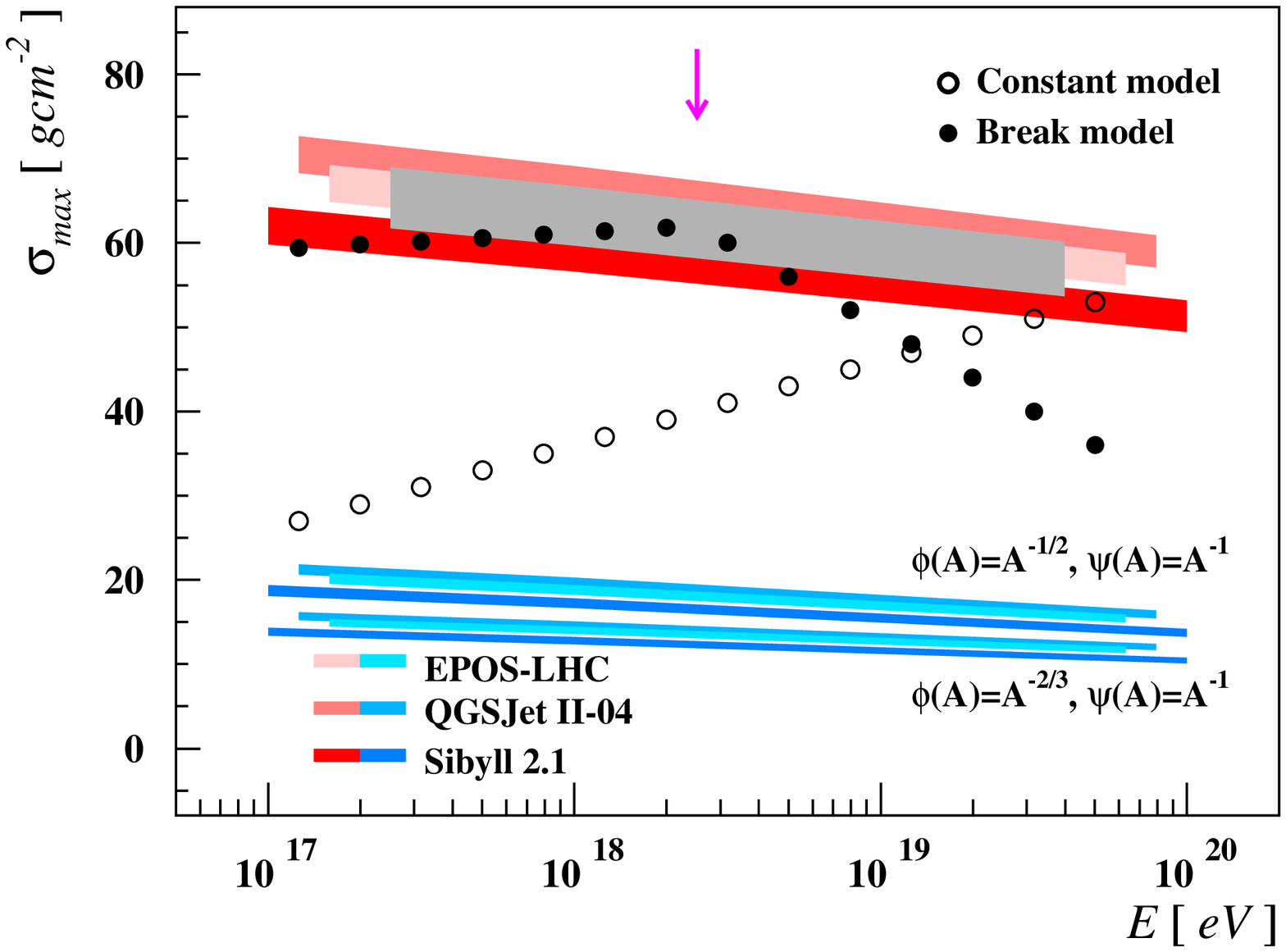}
\caption{\small
Standard deviations of the depth of shower maximum are shown 
as functions of primary energy.
Circles are for the logarithmically increasing standard deviations 
(constant model).
Black full points are for the break model with a break indicated 
by a magenta arrow.  
Bands with red (blue) shade are for $\sxmax$ predictions for 
primary protons (iron nuclei). 
They are based on the estimates of shower fluctuations given in~\app{App01} 
with parameters deduced from CONEX showers generated with primary energies 
around $1 \EeV$ within the indicated models of hadronic interactions. 
Lower (upper) blue bands for iron nuclei are for the mass dependent
terms $\phi(A) = A^{-\frac{2}{3}} (A^{-\frac{1}{2}})$ while $\psi(A) = A^{-1}$, 
see~\app{App01}.
Our parametrization of the square root of the sum of shower variances 
for protons is shown by a gray band with $5 \%$ uncertainties included.
}
\label{F02}
\end{figure}

The shower observables, $\AXmax$ and $\SXmax$, were decomposed 
using different sets of primary particles.
Since it was not feasible to consider all the possible nuclei,
we limited our analysis to primaries representing light, 
intermediate and heavy nuclei. 
We present results with primary protons ($A=1$), and helium 
($A=4$), nitrogen ($A=14$) and iron ($A=56$) nuclei. 
This option is useful for examination of the impact of individual 
components in terms of reducing their number. 
Also other possibilities are briefly mentioned.

The obtained results depend on the quartet of shower parameters, 
namely, $C = \Axmax(E_{0})$, $D$, $\sfro = \sfr(A=1,E_{0})$ and 
$\ssho = \ssh(A=1,E_{0})$, and also on functions describing 
mass dependent fluctuations of the EAS development.  
Given the shower observables, $\AXmax(E)$ and $\SXmax(E)$, 
it turned out that only a certain set of the parameters 
$C$ and $D$ allows to find solutions to the partition problem
in the whole energy range studied.
The impact of other parameters that are associated with shower 
fluctuations was also examined.

With the aim of obtaining useful information from 
the data, we focus on how to exploit the potential of 
the partition concept.
In this study, possible uncertainties of the $\Xmax$ statistics and 
uncertainties of the energy scale are not taken into account. 
The estimates of their impact on resulting solutions can 
be obtained in repeated calculations with input data shifted 
according to these uncertainties.
The ambiguities in resultant distributions with maximum entropy, 
as indicated below, occur herein just through unknown details 
of EAS physics.
They are mainly associated with unknown though acceptable values 
of the parameters $C$ and $D$ that determine properties of proton 
showers.
When presenting our final results of the mass decomposition, 
the values of these parameters are kept near the ranges 
reported by the CR experiments
considering the predictions of the models of hadronic 
interactions~\cite{Ung01,Bar01}.

In a preliminary analysis, we successfully applied the maximum 
entropy method to a number of hypothetical examples.
In order to test this method we assumed different mixtures 
of four or more primary species. 
With each chosen set of primary fractions we generated the two lowest 
order $\Xmax$ moments within the shower model with fixed 
parameters. 
Then, these data were decomposed back to individual components,
as dictated by the partition method. 
All the selected input tasks dominated by very different 
types of primaries were positively identified when seeking 
solutions with four or more components. 
Also predetermined fractions of the lightest species were
reproduced with sufficient accuracy. 
They did not change much when extra particles were added.
In complex ambiguous cases, it turned out, as expected,
that the method of maximum entropy prefers to attribute similar 
occupation probabilities to heavier primaries.

In the following, we present results of two more sophisticated 
examples which provided us with nontrivial solutions.  
A special attention was paid to cases with the energy evolution 
of shower characteristics reminiscent of their measured dependencies. 
In a constant (elongation rate) model, we assumed that the average 
depth of shower maximum and its standard deviation increase linearly 
with the logarithm of the primary energy. 
In the second example, in a break model, we modeled breaks 
in the energy dependence of these statistics.
In both cases, the two lowest order $\Xmax$ moments were 
given for energies ranging from 
$\Log(E/\eV) = 17.1$ to $19.7$ for $14$ values in steps 
of $0.2$.
The input dependencies are visualized in~\figs{F01} and~\figg{F02}.
In these figures, we also show the $\AXmax = \AXmax(E)$ predictions 
of the superposition model and the estimates of shower fluctuations, 
$\sxmax = \sxmax(E)$, derived with the approach of~\app{App01}. 
All these estimates were obtained with the parameters calculated within
the indicated models of hadronic interactions, see~\sct{Sec02}.

\subsection{Constant model}
\label{Sec04a}

The motivation for this example was to show what kind 
of solution is achieved when the input sample moments indicate 
that the transition from a predominantly heavy to light composition 
takes place.
We used the average depth of shower maximum with a constant 
elongation rate.
We assumed that the standard deviation of shower maximum grows 
logarithmically with the primary energy.
Both these shower statistics, $\AXmax(E)$ and $\sxmax(E)$, 
displayed by circles in~\figs{F01} and~\figg{F02}, were chosen 
in such a way that  
\beql{E10}
\frac{\AXmax(E) - X_{0}}{D_{0}} =
\frac{\sxmax(E) - \sigma_{0}}{s_{0}} = 
\Log \left( \frac{E}{E_{0}} \right).
\eeq
Here we used $X_{0} = 673~\gIcmS$ and $\sigma_{0} = 36~\gIcmS$ for the 
$\Xmax$ statistics given at a reference energy of $E_{0} = 1~\EeV$.
The remaining parameters were 
$D_{0} = 80~\gIcmS$ and $s_{0} = 10~\gIcmS$.
\begin{figure}[ht!]
\wse
\includegraphics[width=0.99\linewidth]{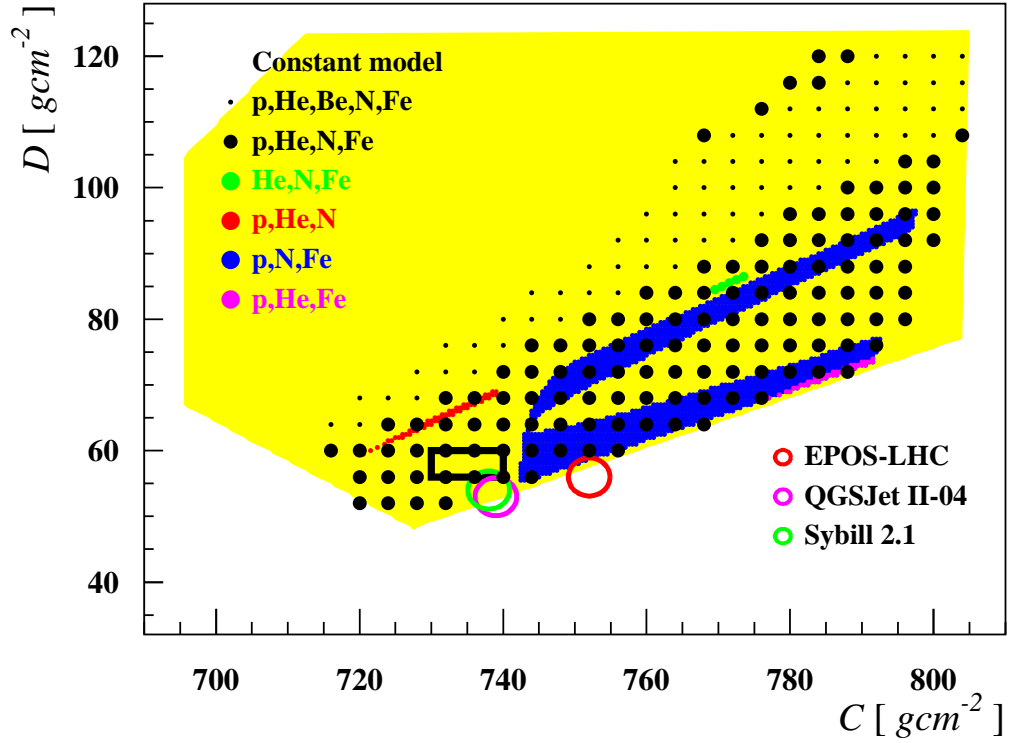}
\caption{\small
$C$--$D$ domains for acceptable solutions of the constant model. 
Yellow area is given by inequalities~(\ref{E05}). 
A domain for acceptable four--component (p,He,N,Fe) solutions to 
the maximum entropy problem, shown by large black points in steps 
of $4~\gIcmS$, is enlarged by five--component (p,He,N,Be,Fe) solutions 
as indicated by small black points.
Green, red, blue and magenta areas show $C$--$D$ domains for 
(He,N,Fe), (p,He,N), (p,N,Fe) and (p,He,Fe) solutions, respectively.
Colored circles indicate the $C$--$D$ ranges for the EPOS-LHC (red), 
QGSJet~II-04 (blue) and Sibyll~2.1 (green) model as obtained 
by fitting their $\AXmax$ values for H, He, N and Fe primary nuclei 
to~\eqa{E01}.
In the partition analysis, we used parameters located inside a black 
rectangle.
}
\label{F03}
\end{figure}
\begin{figure}[ht!]
\wse
\includegraphics[width=0.99\linewidth]{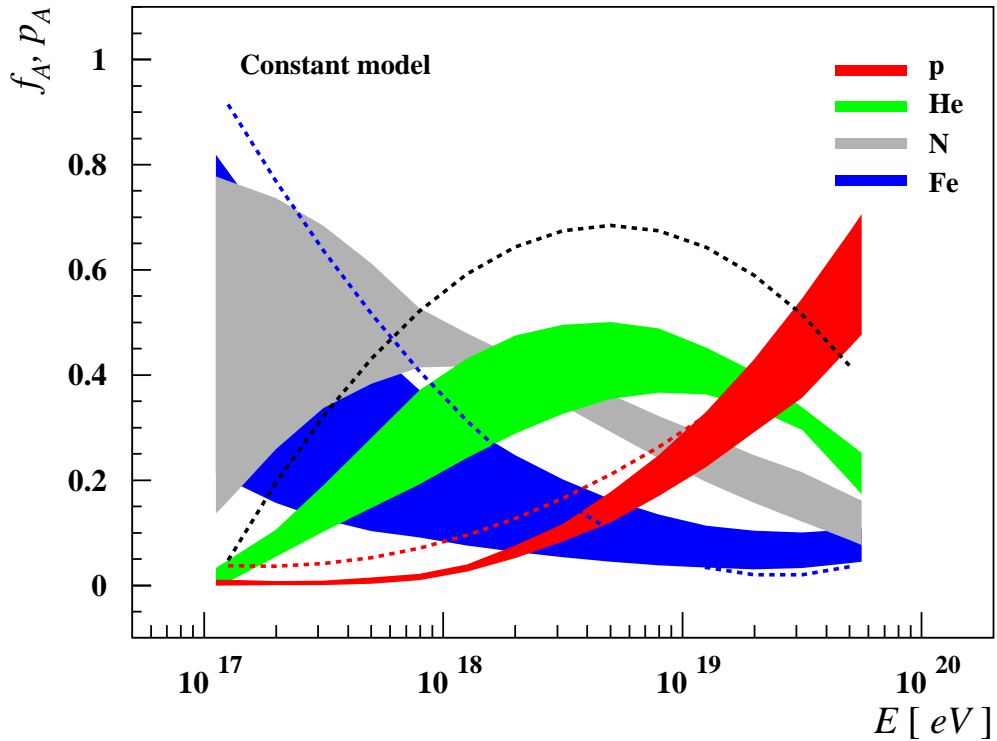}
\caption{\small
Probabilities of (p,He,N,Fe) partition of the constant model 
are depicted as functions of primary energy. 
Red, green, gray and blue bands are for proton, helium, nitrogen 
and iron components, respectively. 
Their widths correspond to the freedom in choosing the model 
parameters $C$ and $D$. 
We used $C \in \langle 730, 740 \rangle \gIcmS$ 
and $D \in \langle 56, 60 \rangle \gIcmS$, 
see the black rectangle in~\fig{F03}. 
Dashed color curves show (p,N,Fe) solution with
$C = 744 \gIcmS$ and $D = 56 \gIcmS$.
}
\label{F04}
\end{figure}

In the first step, we solved the partition problem numerically 
with four or more components leaving the two parameters 
$C$ and $D$ free. 
For the shower fluctuations, we used parametrizations that 
are described in~\app{App01}, see~\eqb{A03} and~\eqc{A04}.
Their parameters were kept constant, i.e. we used
$\sfroo = 46 \gIcmS$ and $\sshoo = 38 \gIcmS$.
For the mass dependent terms we adopted $\phi(A) = A^{-\frac{2}{3}}$ 
and $\psi(A) = A^{-1}$.
We assumed that the variance of the first interaction is energy 
dependent as given in~\eqa{A03} where we set
$\xi(E) = 1 - 0.2~\Log(E/\EeV)$.

Using these assumptions, we searched for domains in the $C$--$D$ 
plane with acceptable maximum entropy solutions.
By this we mean that the input information is compatible with 
the shower model, i.e. the accuracy with which the set of constraints 
written in~\eqb{E08} and~\eqc{E09} is satisfied in the whole considered 
energy range is better than $10^{-2}$. 

Resultant $C$--$D$ domains are visualized in~\fig{F03}.
The yellow area, the part of which is shown in the figure, 
is constructed using inequalities~\eqc{E05} for the primary 
masses $1 \le A \le 56$.
Large black points in~\fig{F03}, shown in steps of $4~\gIcmS$, 
represent possible values of the parameters $C$ and $D$ for which 
four--component solutions were determined.
Acceptable solutions with five components enlarge the space for 
the model parameters as indicated by small black points.
We also show the ranges of the parameters $C$ and $D$ 
that follow from our fits of $\AXmax$ values to the superposition 
ansatz while using CONEX showers initiated by H, He, N and Fe primary 
nuclei within the indicated generators of hadronic interactions.

The trend for the parameters $C$ and $D$ is well visible.
The acceptable four-- or more--component solubility of 
the maximum entropy problem demands larger proton elongation 
rates, $D > 50 \gIcmS$.
Increasing the parameter $C$, the acceptable values for 
the proton elongation rate should be even larger. 
Adding one or more light components ($A = 6-12$) to 
the four--component conjecture, the size of the acceptable region 
is further enlarged towards larger values of the proton elongation 
rate $D$ (small black points in~\fig{F03}).
With the increasing number of intermediate or heavier components
($A = 16-54$), the $C$--$D$ regions for acceptable solutions 
with five or more components remain within the boundaries given 
by the domain for four--component solutions.
In this case we conclude that the most important part of 
the input information is exhausted by four--component 
solutions.  

Different values for the parameters $C$ and $D$ of existing
three--component solutions of the system of~\eqb{E03} and~\eqc{E04} 
are also indicated in~\fig{F03}.
These solutions exist only in the well separated $C$--$D$ regions.
The most of such deterministic solutions is for the triad of 
primary particles (p,N,Fe), see the blue regions in~\fig{F03}.

In the second step, we made an attempt to derive the energy 
evolution of partition probabilities for primary particles 
causing showers with given $\Xmax$ statistics.
We performed the partition analysis with a selected set of 
the parameters $C$ and $D$ that yield acceptable solutions 
to the maximum entropy problem.
Having no other supporting facts we simply explored 
a certain $C$--$D$ region in which the two constraints written 
in~\eqb{E08}~and~\eqc{E09} are satisfactorily fulfilled. 
We focused on small values of the proton elongation rate $D$, 
as observed in CR experiments (see~\sct{Sec02}), and those solutions 
that do not contradict the predictions of the current models 
of hadronic interactions (see also~\figs{F01} and~\figg{F02}).
In making this decision, the whole set of energy dependent 
constraints was considered. 
In particular, the parameters were chosen in the vicinity 
of the predictions of the QGSJet~II-04 and Sibyll~2.1 models, 
as indicated in~\fig{F03}.

A four--component result of the partition analysis that was
obtained with the parameters indicated in the figure caption
(see black rectangle in~\fig{F03} and also middle light 
gray band in~\fig{F01}) is plotted in~\fig{F04}.
In this figure, the decomposition probabilities of primary 
particles derived from the given $\Xmax$ statistics are shown 
as functions of the primary energy.
The widths of the colored bands correspond to aforementioned 
freedom in the parameters $C$ and $D$.
We verified that these bands also comprise four--component 
solutions with the values of the parameters of shower fluctuations 
shifted by $\pm 5 \%$ (see~\fig{F02}) and with different mass 
dependent functions as described in~\app{App01}.
For the sake of comparison, one selected three--component solution 
that does not include helium nuclei (blue region in~\fig{F03}) 
is also shown in~\fig{F04}.

For the sample mean and variance of the depth of shower maximum 
logarithmically growing with the primary energy, the transition 
from heavier to light primaries was identified in the resulting 
probability distributions of primary particles.
A mixture of heavier primaries with very uncertain occupation 
probabilities together with negligible contributions of lighter 
species was obtained at the lowest primary energies.
With the increasing energy, as both the $\Xmax$ statistics 
grow, lighter primaries were found to be responsible for this 
kind of behavior, since the probability of finding them in 
a primary beam increases.
The second lightest primaries (He or N nuclei) play a crucial 
role during the transition from shallow to deep showers accompanied 
by the increase of fluctuations in the depth of shower maximum.
The partition method provides a solution in which a well established 
proton component dominates at the highest energies.
We verified that the course of the transition from heavier 
to light primaries, as found in the four--component solutions, 
remains nearly unchanged with the increasing number of primaries 
included.

\subsection{Break model}
\label{Sec04b}

We also analyzed a set of shower statistics that resembles 
the experimental data that have been collected by the fluorescence detector 
of the Pierre Auger observatory~\cite{Aug01,Aug02,Aug03,Aug04}.
We prepared the input data, $\AXmax(E)$ and $\sxmax(E)$,
with breaks as depicted in~\figs{F01} and~\figg{F02} by black full points. 
For the energy dependence of the average depth of shower maximum 
we chose the elongation rate of $D_{0} = 80~\gIcmS$ for the primary 
energies below $\Log(E/\eV) = 18.4$ and $D_{0}' = 27~\gIcmS$ above 
this energy with the parameter $X_{0} = 708~\gIcmS$, see~\eqa{E10}.
For the energy dependence of the standard deviation of the depth 
of shower maximum (see~\eqa{E10}), we took the value 
$s_{0} = 2~\gIcmS$ for $\Log(E/\eV) < 18.4$, 
$s_{0}' = -20~\gIcmS$ otherwise, and $\sigma_{0} = 61.2~\gIcmS$.

First, we were interested in permissible values for the parameters 
of the proton showers, $C$ and $D$.
Assuming four or more primary species are responsible for 
the energy dependent $\Xmax$ moments, we examined different 
types of maximum entropy solutions.
We used the same parametrization for the shower fluctuations
as given in the previous example in~\sct{Sec04a}.
\begin{figure}[ht!]
\wse
\includegraphics[width=0.99\linewidth]{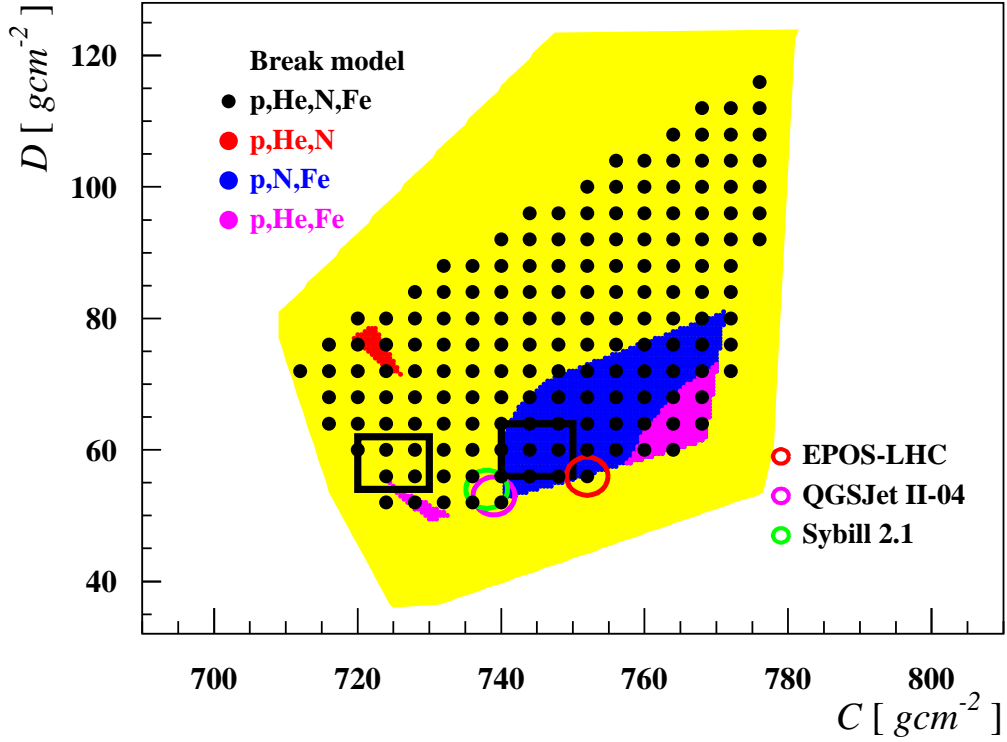}
\caption{\small
$C$--$D$ domains for acceptable solutions of the break model. 
Yellow area is given by inequalities~(\ref{E05}). 
Large black points, shown in steps of $4~\gIcmS$, represent 
acceptable four--component solutions to the maximum entropy 
problem for the (p,He,N,Fe) primary beam. 
Red, blue and magenta areas show $C$--$D$ regions for (p,He,N), 
(p,N,Fe) and (p,He,Fe) solutions, respectively.
The region of (p,N,Fe) solutions (shown in blue) is partially 
overlapped with the (p,He,Fe) region (drawn in magenta).
Colored circles indicate the ranges of $C$ and $D$ for 
the EPOS-LHC (red), QGSJet~II-04 (blue) and Sibyll~2.1 (green) 
model. 
In the examples shown in~\figs{F06} and~\figg{F07}, 
we used parameters lying inside the left and right black 
rectangle, respectively.
}
\label{F05}
\end{figure}
\begin{figure}[ht!]
\wse
\includegraphics[width=0.99\linewidth]{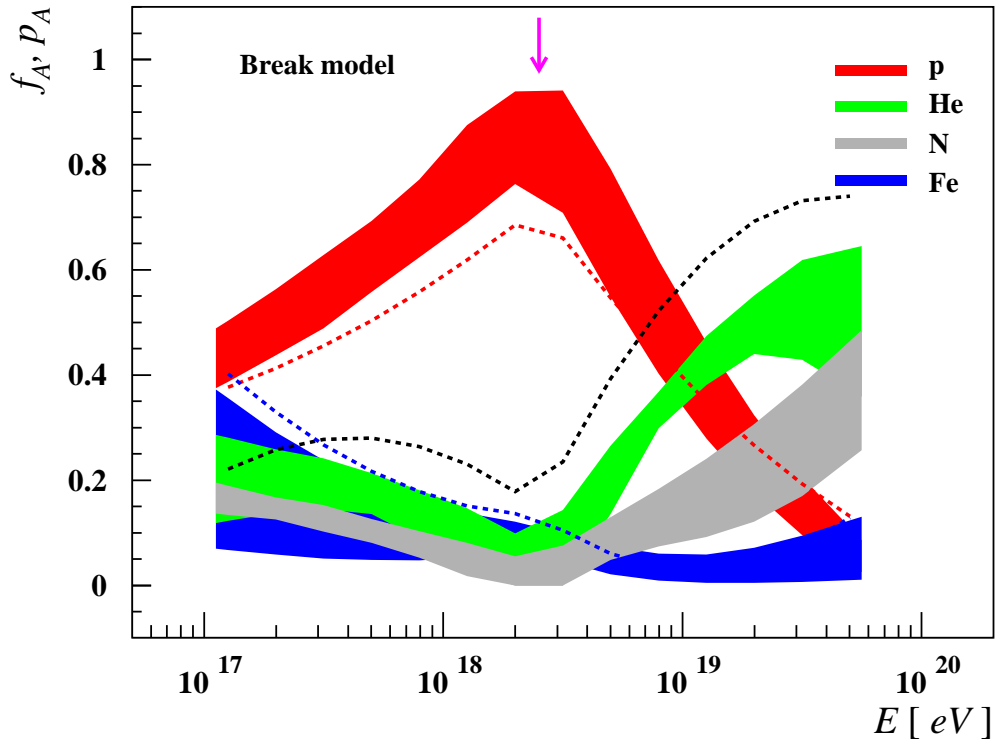}
\caption{\small
Probabilities of (p,He,N,Fe) partition for the break model 
are depicted as functions of primary energy, see also caption 
to~\fig{F04}.
We used  
$C \in \langle 720, 730 \rangle \gIcmS$ 
and $D \in \langle 54, 62 \rangle \gIcmS$,
see the left black rectangle in~\fig{F05}. 
Dashed color curves show (p,N,Fe) solutions with 
$C = 740 \gIcmS$ and $D = 56 \gIcmS$. 
}
\label{F06}
\end{figure}
\begin{figure}[ht!]
\wse
\includegraphics[width=0.99\linewidth]{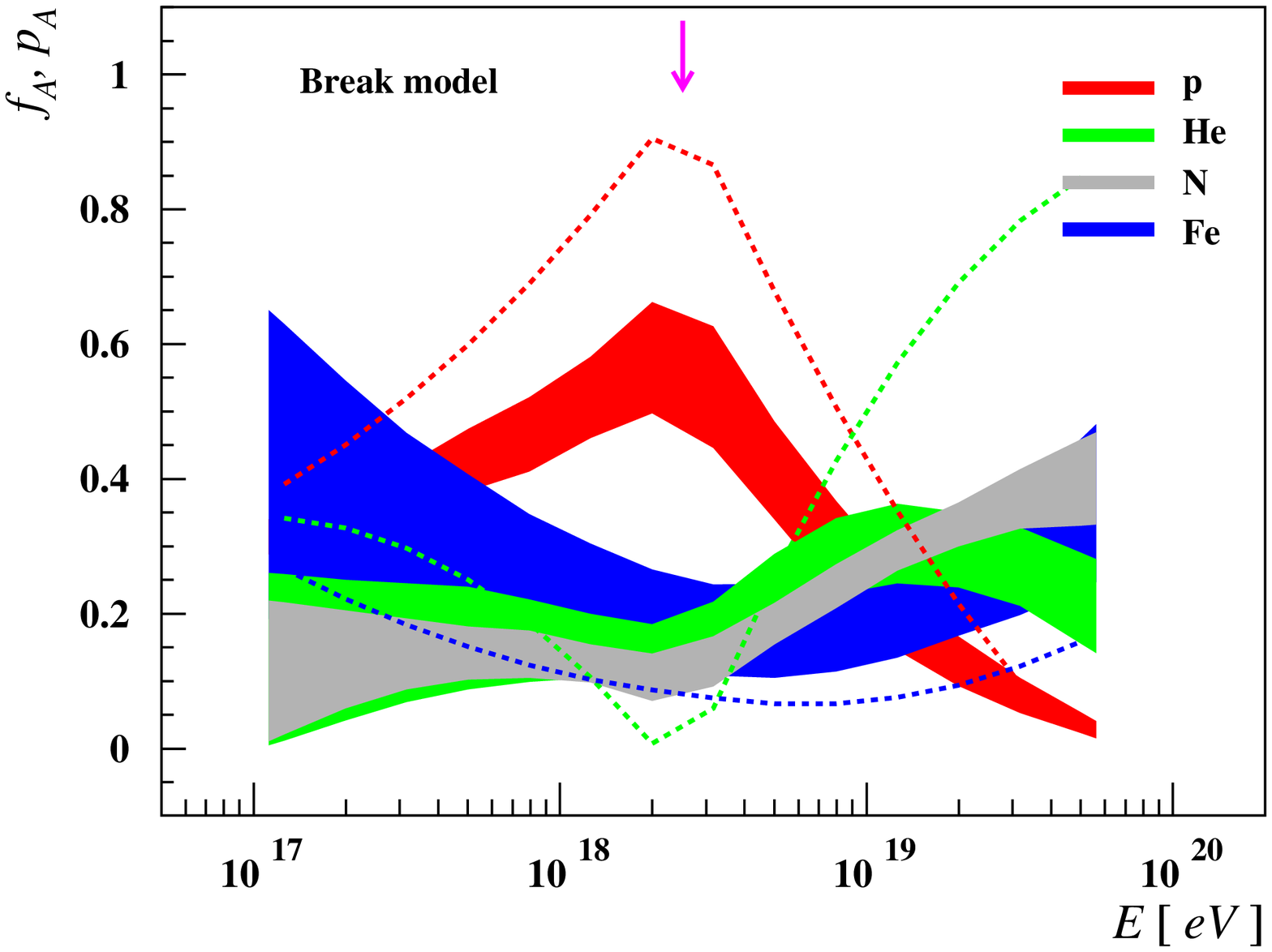}
\caption{\small
Probabilities of (p,He,N,Fe) partition for the break model 
are depicted as functions of primary energy, see also caption 
to~\fig{F04}.
They were obtained with the parameters   
$C \in \langle 740, 750 \rangle \gIcmS$ 
and $D \in \langle 56, 64 \rangle \gIcmS$,
see the right black rectangle in~\fig{F05}. 
Dashed color curves show (p,He,Fe) solutions with 
$C = 724 \gIcmS$ and $D = 56 \gIcmS$.
}
\label{F07}
\end{figure}

Resultant domains of the parameters $C$ and $D$ that provide 
solutions in the whole energy range are shown in~\fig{F05}.
The yellow area, the part of which is depicted in the figure,
corresponds to inequalities~(\ref{E05}) for the masses 
$1 \le A \le 56$.
The model parameters of acceptable four--component solutions 
were determined in steps of $4~\gIcmS$, as shown by large 
black points.
With further increase of the number of primary components above four, 
the $C$--$D$ regions giving acceptable solutions stay within 
the boundaries of the four--component $C$--$D$ domain.
Also, the $C$--$D$ regions for different types of three--component 
solutions are indicated in~\fig{F05}.
Colored circles show the ranges of relevant parameters estimated 
with the help of the models of hadronic interactions.

In this example, we learned that the input information on the energy 
evolution of the two lowest order $\Xmax$ moments is well described 
by four--component solutions to the maximum entropy problem. 
A range of the model parameters $C$ and $D$ providing these 
solutions is not in conflict with their current estimations~\cite{Ung01,Bar01}.

In order that three--component solutions exist for all considered 
energies, the proton elongation rate $D > 50 \gIcmS$ and even larger 
value ($D > 70 \gIcmS$) is required if three--component solutions 
without iron nuclei are constructed, see the red region in~\fig{F05}.
We found that the proton component cannot be missed in any case.
There is a large domain of the parameters $C$ and $D$ providing 
three--component solutions without primary helium, see the blue 
region in~\fig{F05} that is partially overlapped with the $C$--$D$ 
domain of the (p,He,Fe) solutions shown in magenta.

Finally, we were concerned with the probability distributions 
of primary species as they can be derived in the decomposition 
analysis using the input information about the two lowest order 
$\Xmax$ moments.
We carried out the partition analysis with the parameters 
$C$ and $D$ that yield acceptable solutions to the maximum 
entropy problem. 
In order to illustrate the impact of the parameter selection, 
we examined in this case two different $C$--$D$ domains.
We chose one set of parameters with small reference values of 
the mean depth of shower maximum for protons and the second set 
located near the predictions of the models of hadronic interactions, 
see two black rectangles in~\fig{F05}.

A resultant four--component partition is shown as a function 
of the primary energy in~\fig{F06}. 
It was obtained with the parameters indicated in the figure caption
(see left black rectangle in~\fig{F03} and also lower dark gray 
band in~\fig{F01}).
Also a three--component partition that does not account for a fraction 
of helium (blue region in~\fig{F05}) is presented in~\fig{F06}.
In~\fig{F07}, a four--component partition to the maximum entropy problem 
obtained for a different set of the model parameters
(see right black rectangle in~\fig{F03} and also upper 
dark gray band in~\fig{F01})
is shown together with another three--component solution 
to~\eqb{E03} and~\eqc{E04}, in this case without nitrogen nuclei.
The breaks modeled in the energy dependencies of the two $\Xmax$
statistics are well visible in the partition probabilities of all 
these solutions.

Within the proposed method, we succeeded in the description of 
energy dependencies of the two lowest order $\Xmax$ moments.
The resultant evolution of the probability distribution of 
particles in the primary beam can be interpreted as follows.
Apart from primary protons, second lightest primaries (He or N nuclei) 
were found to play an important role in the presented decomposition.
The probability of the proton presence in the primary beam initially 
grows and then declines rapidly with the primary energy once it 
reaches its maximum value near the modeled break point at about 
$\Log(E/\eV) = 18.3$. 
Around this energy the remaining components are very likely to be 
drastically reduced.
The proton dominance near the break point is diminished with 
the increasing value of the parameter $C$, while the role of 
the heaviest component (Fe nuclei) grows, compare~\fig{F06} with~\fig{F07}. 
After the break point, the modeled transition from shallow to
deep showers with a fall of fluctuations in the depth 
of shower maximum does not demand any substantial proton component.
Instead, particles with intermediate masses or a complex mixture 
of intermediate and heavier nuclei were found to take over the main role.
Adding one or more primary species, the typical features of this 
partition are preserved. 

\section{Conclusions}
\label{Sec05}

We used the well--reasoned method based on the maximum entropy 
principle to derive the partition of primary cosmic ray particles 
from the characteristics of the longitudinal development of extensive 
air showers that they initiated.
A set of the first and second order $\Xmax$ moments is used as 
the input data.
The proposed approach combines the superposition ansatz and 
multiplication characteristics of air showers.
As input parameters, the method needs the values of the mean 
depth of shower maximum for protons and of the elongation rate 
for protons, both at a chosen reference energy.
The mass dependent terms associated with shower fluctuations, 
as inferred from experimental data, were proved to be applicable 
in the search for the primary mass composition. 

We showed that simple assumptions about the properties of extensive
air showers make the partition analysis suitable for exploring energy 
dependent changes in the composition of the primary beam of cosmic 
particles.
In particular, we presented solutions to the partition problem for 
two hypothetical sets of the two lowest order $\Xmax$ moments. 
It was shown that these data can be satisfactorily described by assuming 
that four groups of primary species are present in the primary beam.
The partition method allowed us to partially reduce the parameter 
space within the model adopted for the shower development.
With a set of parameters selected in agreement with 
experiments and model predictions we were able to assess 
specific solutions.
While the role of the lightest primaries is indisputable,  
heavier particles were identified with more uncertainty.
Our analysis indicates that primaries of intermediate masses 
are required when searching for reasonable solutions.

The proposed partition scheme constitutes a very simple 
way to extract undistorted information about the decomposition 
of the primary cosmic rays into individual species from 
the measured data.
It relies on well known statistical arguments which delivers 
a special interpretation to results.
Deduced in the maximum entropy method, the obtained partitions 
and subsequent conclusions have probabilistic nature by definition.
In the analysis of shower observables, the resultant probability 
distributions of primary particles are least affected with regard 
to missing information, while respecting our knowledge of shower 
physics inserted into the calculations. 
In this sense, the selective ability of maximum entropy 
may help in the classification of available models of hadronic 
interactions.

\appendix

\section{Shower variances}
\label{App01}
\setcounter{figure}{0}

In our method, the mean depth of shower maximum for
primary protons with an energy $E$ is assumed to be~\cite{Ung01}
\beql{A01}
\Axmax(E) \approx \lambda(E) + X~\ln 
\left( \frac{\kappa E}{2 M \epsilon} \right),
\eeq
where $\lambda = \lambda(E)$ is the mean interaction length 
for inelastic $p$--air collisions, $X \approx 37~\gIcmS$ is 
the radiation length in air and $\epsilon \approx 84~\MeV$ 
denotes the critical energy in air. 
Both the elasticity of the first interaction, $\kappa = \kappa(E)$, 
and its multiplicity, $M = M(E)$, are energy dependent.
The relationship written in~\eqa{A01} is well documented 
within the Heitler model extended to hadronic showers~\cite{Ung01}.

For the variance of the depth of the first interaction we employed
the measured $p$--air cross section at a center of mass energy 
of $\sqrt{s} = 57~\TeV$~\cite{Aug16}.
Relying upon a smooth extrapolation from accelerator measurements, 
and in agreement with model predictions, we used for its 
parametrization
\beql{A02}
\Sigma_{\rm p-Air} \approx \left[ 500 + 50~\Log(E/\EeV) \right] \mbb.
\eeq
Within a naive model, the variance of the depth of the first 
interaction of a primary nucleus with $A$ nucleons colliding 
with the air target is then approximately 
\beql{A03}
\sfr = \sfr(A,E) \approx \phi(A) \xi(E) \sfro, 
\eeq
where $\phi = \phi(A)$ denotes a mass dependent term and 
$\xi = \xi(E)$ is a general dependence of the variance 
on the primary energy. 
A value of the square root of the variance of the depth
of the first interaction caused by primary protons at 
a reference energy of $E_{0} = 1~\EeV$, 
$\sfroo = \lambda(E_{0}) \approx 46~\gIcmS$, 
as well as a constant in the energy dependent function 
$\xi(E) \approx 1 - 0.2~\Log(E/\EeV)$, 
were deduced from the parametrization given in~\eqa{A02}.

The mass dependent term in~\eqa{A03}, $\phi = \phi(A)$, accounts 
for the details of the first interaction given by individual 
nucleon--nucleon interactions and subsequent nuclear 
fragmentation~\cite{Ung01}. 
Averaging over all available numbers of simultaneously interacting 
nucleons accompanied by the remaining number of free spectator 
nucleons, we obtained for the variance of the depth of the first 
interaction 
$\phi_{1}(A) =  
\frac{1}{3} + \frac{1}{A} - \frac{1}{3A^{2}} > A^{-\frac{1}{2}}$.
Since $\phi_{0}(A) = A^{-1}$ is expected for $A$ free nucleons, 
we chose the mass dependent function 
$\phi(A) = A^{-\alpha}$ with a constant index 
$\alpha = \frac{2}{3}$, 
i.e. $\phi_{0} < \phi < \phi_{1}$ for any $A$.
We also examined values of $\alpha$ ranging from $\frac{1}{3}$ 
up to~$1$. 
These values yielded slightly different results, with deviations 
below uncertainties that are due to the ambiguity of other shower 
parameters.

The variance of the depth of shower maximum in the subsequent 
shower development is given by
\beql{A04}
\ssh = \ssh(A,E) \approx \psi(A) \ssho,
\eeq
where the function $\psi = \psi(A)$ is responsible for the mass 
effects. 
The parameter $\sshoo \approx 38~\gIcmS$, the square root of 
the shower variance for primary protons, was estimated from 
the standard deviation of the depth of shower maximum 
$\sxmax \approx 60~\gIcmS$ measured at about 
$1~\EeV$~\cite{Aug01,Aug02,Aug03,Aug04}, when mostly primary 
protons are assumed to cause such showers.
Experimentally, the standard deviation of the depth of shower maximum 
declines with the increasing energy showing that heavier primaries 
with a narrower mass distribution are 
responsible~\cite{Aug04,Ung01,Bar01,Aug07}.
Hence, the square root of the shower variance for primary protons
has a lower bound somewhere below the indicated value when an 
unknown primary composition is observed at the reference energy.

The mass dependent term of the shower variance, $\psi = \psi(A)$, 
is given by fluctuations in multiplicity $M$ and elasticity 
$\kappa$ of the first (or main) interaction. 
Assuming a simple concept introduced in~\eqa{A01}, the corresponding 
shower variance caused by primary protons is
$\ssho = \smuo + \skao$ where
\beql{A05}
\smuo \approx X^{2} \frac{\smu}{M^{2}}, \qqc
\skao \approx X^{2} \frac{\ska}{\kappa^{2}}.
\eeq 
Here $\smu$ and $\ska$ denote, respectively, the variances 
in multiplicity and elasticity associated with the main 
interaction. 
In a naive superposition model~\cite{Ung01}, the variance 
of the total multiplicity of $k$ out of $A$ nucleons 
participating in the main interaction with a multiplicity 
$\ovl{M}$ averaged over all projectile nucleons
is given by $\sigma^{2}(k \ovl{M}) = k^{2} A^{-1} \smu$. 
In a similar manner, one gets 
$\sigma^{2}(\ovl{\kappa}) = A^{-1} \ska$ 
for an average elasticity $\ovl{\kappa}$
taken over all projectile nucleons.
Hence, we used $\psi(A) = A^{-1}$ in this approach.
\begin{figure}[ht!]
\wse
\includegraphics[width=0.99\linewidth]{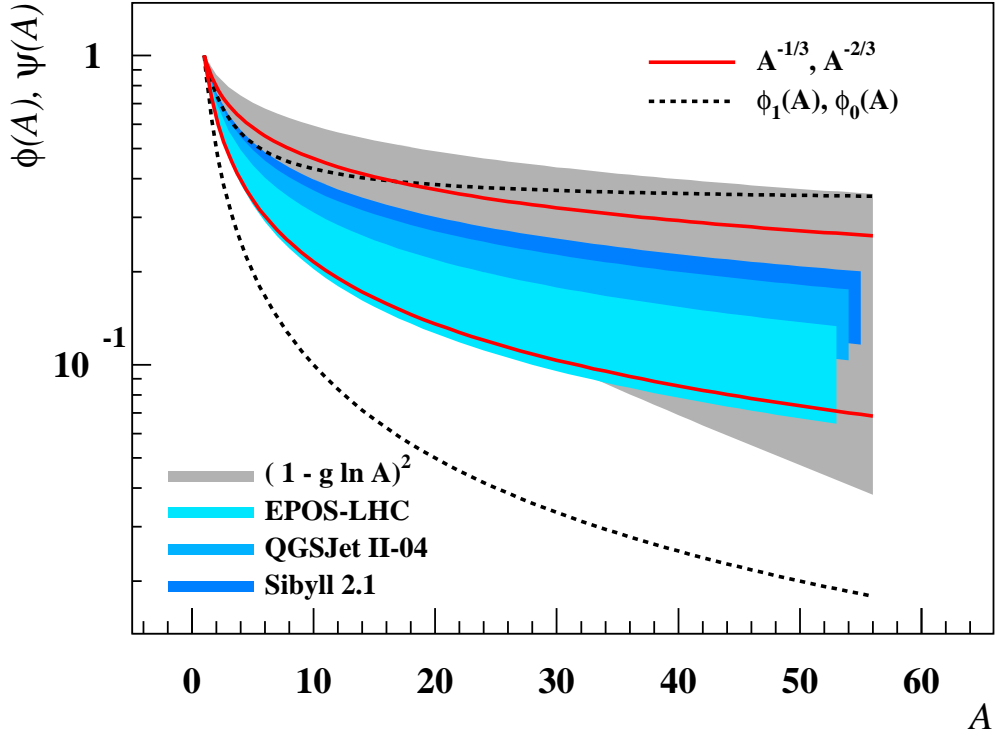}
\caption{\small
Parametrization of shower variances.
Black dashed curves show, respectively, an average value for 
the $A$--dependent term of the variance of the depth of the first 
interaction with one cluster of nucleons, $\phi_{1}(A)$ (upper
curve), or with free nucleons, $\phi_{0}(A)$ (lower curve). 
The mass dependent terms $\phi(A) = A^{-\frac{1}{3}}$ (upper curve) 
and $\phi(A) = A^{-\frac{2}{3}}$ (lower curve) are shown in red.
Gray band is for the parametrization of the shower variances 
taken from Ref.~\cite{Lin01}.
Bands in blue shade represent our parametrizations obtained with 
CONEX showers using the EPOS-LHC, QGSJet~II-04 and Sibyll~2.1 models.
}
\label{F08}
\end{figure}

In order to get a better overview of the problem of the shower 
fluctuations, we examined the logarithmic parametrization 
of the total shower variance as suggested in Ref.~\cite{Lin01}. 
Specifically, we took 
$\phi(A) = \psi(A) \approx \left( 1 - g \ln A \right)^{2}$,
where the parameter $g = 0.15 \pm 0.05$~\cite{Lin01}.
The energy dependence for the variance of the depth of the first 
interaction was included, see~\eqa{A03}.
Since air shower simulations are known to predict larger shower 
variances for heavy primaries than expected from naive 
considerations, we also examined less steep dependencies on 
the primary mass, namely, $\phi(A) = \psi(A) = A^{-\frac{1}{2}}$.
This choice was supported by overall shower fluctuations 
that were derived using a set of CONEX showers~\cite{Ber01} generated 
with energies around $1 \EeV$ for H, He, N and Fe primary nuclei 
within the EPOS-LHC~\cite{Wer01,Wer02}, QGSJet~II-04~\cite{Ost01} 
and Sibyll~2.1~\cite{Ahn01} models. 
In these simulations, we received mass dependent terms 
$\phi(A) = \psi(A) \approx A^{-\alpha}$ where 
$\alpha \approx 0.58, 0.49$ and $0.46$, respectively, with 
uncertainties about $15\%$.
In summary, dealing with each of the above assumptions, 
we obtained similar results.

Different mass dependent functions used for the shower variances 
are depicted in~\fig{F08}. 
While these variances are little dependent on the primary mass 
for $A > 20$, lighter nuclei ($A < 10$) give very different 
contributions.
Based on this approximation that pursue the current knowledge 
of the shower physics, medium and heavier nuclei can hardly 
be distinguished on the basis of shower fluctuations.

\section{Maximum entropy formalism}
\label{App02}

Let us assume that the quantity $A$ is capable to take $n$ discrete
values $A=1,\dots,n$.
Corresponding probabilities $\pa$ are not known, however. 
Only a set of $r$ expectation values of the functions 
$F_{i}(A)$, $i=1,\dots,r$, $r < n$, is measured.
For setting up a probability distribution which satisfies 
the given data, the least biased estimate possible on the basis 
of partial knowledge is used.
The underlying information--theoretic principle, known as the maximum 
entropy principle, was originally introduced in the context of
statistical physics~\cite{Jay01}.
For its use in statistics see e.g. Ref.~\cite{Rao01}.

The overriding principle of this procedure matches an intuition 
of how maximally unbiased estimates should be achieved in the absence
of detailed information about investigated phenomena.  
Here, Shannon entropy~\cite{Jay01,Pre01} 
\beql{B01}
S = - k \Ssum{A=1}{n} \pa \ln \pa,
\eeq
where $k$ is a positive constant, is used as an information 
measure of the amount of uncertainty in the probability 
distribution $\pa$ of the quantity $A$.
This distribution is determined as the one that maximizes 
entropy $S$ in~\eqa{B01} subject to $r$ constraints, 
$F_{i}(A)$, $i=1,\dots,r$, given their averages that represent
whatever experimental information one has, i.e. 
\beql{B02}
\langle F_{i} \rangle = \Ssum{A=1}{n} \pa F_{i}(A), \qqa 
i=1,\dots,r, 
\eeq 
subject to the normalization condition $\Ssum{A=1}{n} \pa = 1$.
Then, the resultant distribution describes what we know about 
the quantity $A$ from experiment without assuming anything else 
about the data~\cite{Jay01}. 

In making inferences on the basis of partial information, 
the maximum entropy probability distribution that maximizes 
Shannon entropy in~\eqa{B01} subject to the experimental 
constraints written in~\eqa{B02} is given by~\cite{Jay01,Pre01}
\beql{B03}
\pa = Z^{-1}
e^{-\left[ \lambda_{1} F_{1}(A) + \dots + \lambda_{r} F_{r}(A) \right]}, 
\eeq
with the partition function written
\beql{B04}
Z(\lambda_{1}, \dots, \lambda_{r}) = 
\Ssum{A=1}{n} 
e^{-\left[ \lambda_{1} F_{1}(A) + \dots + \lambda_{r} F_{r}(A) \right]},
\eeq
and with Lagrangian multipliers $\lambda_{i}$, $i=1,\dots,r$, 
to be determined. 
Then, based on Shannon's information measure, the resultant 
probability distribution $\pa$ of the quantity $A$ obtained 
in this process is spread out as widely as possible without 
contradicting the available experimental information represented 
by the average values of the constraints. 
For further details see e.g. Ref.~\cite{Pre01} and 
references therein.

\vspace*{1.3cm}
{\bf Acknowledgment:}
We would like to acknowledge many useful discussions with
our colleagues of the Pierre Auger Collaboration.
We thank to two unknown reviewers that help us with 
the presentation of this study.
This work was supported by the Czech Science Foundation grant 14-17501S.
The research of J.N. was supported by the Czech Science Foundation 
under project GACR~P103/12/G084.
\vspace*{1.3cm}






\end{document}